\documentclass[useAMS,usenatbib,fleqn]{mnras}
\usepackage{newtxtext,newtxmath}
\usepackage[T1]{fontenc}
\usepackage{ae,aecompl}
\usepackage{graphicx,xspace}   
\usepackage{amsmath}    
\usepackage{amssymb,times}    

\def\boo{Bo\"{o}tes~I\xspace}
\def\Teff{$T_{\rm eff}$\xspace}
\def\lgg{log\,$g$\xspace}
\newcommand{\eps}[1]{\log\varepsilon_{\rm #1}}
\def\kms{km\,s$^{-1}$\xspace}
\def\paperI{\citet[][]{2017A&A...604A.129M}\xspace}
\def\paperII{\citet[][]{2017A&A...608A..89M}\xspace}
\def\afe{[$\alpha$/Fe]\xspace}
\def\feh{[Fe/H]\xspace}
\def\kH{$S_{\!\rm H}$}    

\def\physscr{Phys.~Scr.}   

\renewcommand{\ion}[2]{#1~{\sc{\romannumeral #2}}}

\begin{document}
	
\title[Contribution of SNIa to the chemical tnrichment of the UFD galaxy \boo]{Contribution of type Ia supernovae to the chemical enrichment of the ultra-faint dwarf galaxy \boo}
	
\author[Yu.V.Pakhomov et al.]{
 \parbox{\textwidth}{
			Yu.~V.~Pakhomov$^{1}$\footnote{Email: pakhomov@inasan.ru},
            L.~I.~Mashonkina$^{1,2}$,
            T.~M.~Sitnova$^{1}$,
            P. Jablonka$^{3,4}$            
		}\\
		\\
		$^1$Institute of Astronomy, Russian Academy of Sciences, Pyatnitskaya 48, 119017, Moscow, Russia\\
		$^2$Herzen State Pedagogical University, Kazanskaya ul. 5, St. Petersburg, 191186 Russia		\\
		$^3$Laboratoire d'Astrophysique, Ecole Polytechnique F\'{e}d\'{e}rale de Lausanne (EPFL), Observatoire de Sauverny, CH-1290 Versoix, Switzerland\\
		$^4$GEPI, Observatoire de Paris, CNRS, Universit\'{e} Paris Diderot, F-92125 Meudon Cedex, France
}

\journal{ISSN 1063-7737, Astronomy Letters, 2019, \textbf{45}, 259--275 \\
\hfill DOI: 10.1134/S1063773719050050}

\makeatletter
\def\ps@plain{%
  \let\@mkboth\@gobbletwo
  \renewcommand{\@oddhead}{\Large\hfill{\it\@shorttitle}\hspace{1.5em} \rm\@ddell\thepage}
  \renewcommand{\@evenhead}{\Large\@ddell\thepage\hspace{1.5em}\it\@shortauthor\hfill}%
  \renewcommand{\@evenfoot}{\hfil\small Astronomy Letters, 2019, \textbf{45}, 259--275}%
  \renewcommand{\@oddfoot}{\small Astronomy Letters, 2019, \textbf{45}, 259--275\hfil}%
}
\pagestyle{plain}
\makeatother
	
\date{\noindent
ISSN 1063-7737, DOI: 10.1134/S1063773719050050\\
Astronomy Letters, 2019, Vol. 45, No. 5, pp. 259--275.
}
	
\pubyear{2019}
	
\maketitle
\label{firstpage}

\begin{abstract}
For three stars in the ultra-faint dwarf (UFD) galaxy \boo we have determined the atmospheric parameters, performed a new reduction of high-resolution spectra from the Subaru archive, and derived the abundances of eight chemical elements without using the LTE assumption. As a result, among the galaxies of its class \boo now has the largest sample of stars (11) with a homogeneous set of atmospheric parameters and chemical abundances, and this makes it the most promising for studying the chemical evolution of UFD galaxies. We show that in the range $-3<$ \feh $<-2$ for each of the three $\alpha$--process elements, magnesium, calcium, and titanium, a transition from their overabundance relative to iron with \afe$\approx$0.3 to the solar \afe ratio occurs. This most likely suggests the commenced production of iron in type Ia supernovae. The behaviour of the carbon, sodium, nickel, and barium abundances does not differ from that in more massive galaxies, our Galaxy and classical dwarf spheroidal galaxies.
\end{abstract}

\section*{Introduction}

\boo, one of the ultra-faint dwarf (UFD) spheroidal galaxy satellites of the Milky Way, was discovered by \citet{2006ApJ...647L.111B} when analyzing the Sloan Digital Sky Survey (SDSS) catalogue \citep{2000AJ....120.1579Y}. It is located at a distance of $\sim$60~kpc and has a radius of $\sim$200~pc. Its absolute magnitude is $M_V\approx -5.8$~mag; this is one of the faintest galaxies. Based on data from \cite{2011ApJ...736..146K}, \cite{2012AJ....144....4M} estimated the stellar mass $M_{\star}=0.029\times10^6 M_{\sun}$ and the dynamical mass $M_{dyn}=0.81\times10^6 M_{\sun}$, i.e., \boo is one of the lowest-mass dwarf galaxies.

The turnoff point, the red giant branch, and the developed horizontal branch, suggesting the predominance of an old stellar population in \boo, are distinguished on the color--magnitude diagram. The absence of star formation at the present epoch is also confirmed by the dearth of gas \citep{2007MNRAS.375L..41B}. 

\begin{table*}
	\centering
	\caption{The list of investigated stars in the galaxy \boo}
	\label{tab:obs} 
	\small
	\begin{tabular}{cccccccc}
		\hline\hline
		Star & RA          & Dec         & $m_V$ & Exposure time & S/N   & Date & $\varv_{rad}$\\
		&\multicolumn{2}{c}{(J2000.0)}&     & h & 5000/6000~\AA &      & \kms\\
		\hline
		009 & 13:59:48.81 & +14:19:42.9 & 17.48 & 5   & 18/29 & 2010/05/15\;~\;~\; & 104.9$\pm$0.8\\
		094 & 14:00:31.50 & +14:34:03.6 & 17.04 & 2.4 & 14/22 & 2009/05/16\;~\;~\; & \;~95.0$\pm$0.8 \\ 
		117 & 14:00:10.49 & +14:31:45.5 & 17.79 & 4.5 & 11/17 & 2010/05/16\;~\;~\; & 100.3$\pm$0.8 \\
		121 & 14:00:36.52 & +14:39:27.3 & 17.47 & 3   & 16/24 & 2009/05/16\;~\;~\; & 105.7$\pm$0.8\\
		911 & 14:00:01.07 & +14:36:51.5 & 17.52 & 4   & 17/26 & 2009/05/15-16      & 102.3$\pm$0.8\\
		\hline
	\end{tabular}
\end{table*}

The chemical evolution of \boo has been studied in a number of works. Based on low-resolution ($R\simeq5\,000$) spectra and using only the Ca~II K line in 16~stars, \cite{2008ApJ...689L.113N} were the first to establish the metallicity range $\Delta$ \feh $\approx$1.7~dex and the lower bound \feh$=-3.4$. Using high-resolution ($R\approx35\,000$) HIRES Keck~I spectra for seven stars, \cite{2009A&A...508L...1F} determined the mean metallicity \feh$=-2.3$ and showed that six of the seven stars have close and typical calcium and magnesium overabundances relative to iron for metal-poor stars, suggesting the dominance of type II supernovae (SNe II) in the chemical enrichment and complete mixing of nucleosynthesis products. It should be noted that although the spectra of these stars were taken with the Keck~I telescope, the signal-to-noise ($S/N$) ratio was low, which led to large errors in determining the abundances. For example, for the star Boo-127 \cite{2009A&A...508L...1F} detected an anomalously high [Mg/Ca] abundance ratio, but this was not confirmed by subsequent studies based on higher-quality spectra \citep{2013ApJ...763...61G, 2014A&A...562A.146I, 2016ApJ...826..110F}. Based on the FLAMES/UVES VLT2 spectra taken for seven stars, four of which were common to the sample of \cite{2009A&A...508L...1F}, \cite{2013ApJ...763...61G} confirmed that the abundances of chemical elements evolved in this low-mass system. For most elemental ratios their dependence on \feh coincides with that observed in the same metallicity range in the Galactic halo. However, \cite{2013ApJ...763...61G} suspected a slope of the \afe--\feh dependence for the $\alpha$-process elements, which may be indicative of prolonged star formation in \boo. \cite{2014A&A...562A.146I} took spectra with the Subaru telescope for six stars, five of which were common to the sample of \cite{2009A&A...508L...1F}, and showed the elemental ratios for stars in this galaxy to be highly homogeneous. They pointed out a constancy of the [Mg/Fe] and [Ca/Fe] ratios, but with a possible decrease with increasing \feh. This was interpreted as a contribution of type Ia supernovae (SNe~Ia) to the production of iron.

The observed behavior of \afe as a function of metallicity provides crucial information about the chemical evolution of the galaxy. From one to seven stars were studied in each of the papers devoted to \boo. If the results of various determinations are gathered together, then we obtain a highly inhomogeneous data set, because different methods of determining the stellar atmosphere parameters, different grids of model atmospheres, different codes, and even different atomic parameters of spectral lines were used in different papers. \cite{2017A&A...604A.129M, 2017A&A...608A..89M} set the goal to revise the available published data on the abundances of chemical elements for extremely metal-poor stars (\feh$<-2$) in dwarf galaxies, including \boo, by redetermining the atmospheric parameters using unified methods and by analyzing high-resolution ($R>20\,000$) spectra using unified methods in which the assumption about local thermodynamic equilibrium (LTE) is abandoned. It should be noted that all of the published data for stars in dwarf galaxies were obtained within LTE. For \boo eight stars from \citet{2013ApJ...763...61G} and \citet{2016ApJ...826..110F} were included in the sample. \cite{2017A&A...608A..89M} concluded that in the range \feh $\le -2.7$  stars in \boo exhibit the same overabundance of $\alpha$-process elements relative to iron, [$\alpha$/Fe] $\simeq$ 0.3, as in the Galactic halo, but for a group of stars with \feh $\approx -2$ the \afe ratio is solar, which can be explained by the already robust production of iron in SNe Ia. The absence of intermediate-metallicity stars, $-2.7<$ \feh $<-2$, did not allow the epoch of appearance of the first SNe Ia in \boo to be established, and this would be very important for understanding the star formation processes in such small galaxies. High-resolution spectra are available for three more stars in \boo. They were taken with the Subaru telescope and were analyzed by  \cite{2014A&A...562A.146I}. However, the Subaru archive was opened only in late 2017 and, therefore, these three stars were not included in the sample by \cite{2017A&A...604A.129M, 2017A&A...608A..89M}. The goal of this paper is to expand the sample of stars in \boo with homogeneous and accurate data on the abundances of chemical elements by analyzing the spectra from the Subaru archive and to refine the \afe--\feh dependence. We use the same methods as those in \cite{2017A&A...604A.129M, 2017A&A...608A..89M}. The reduction of the spectra from the Subaru archive is described in Section~\ref{Sect:obs}; the atmospheric parameters are determined in Section~\ref{Sect:Param}. We determine the abundances of chemical elements and analyze the elemental ratios in Sections~\ref{Sect:Abund} and \ref{Sect:discussion}. Finally, the conclusions are formulated in Section~\ref{Sect:conc}.

\section{The sample of stars and observational data}
\label{Sect:obs}

Three stars, Boo-009, Boo-121, and Boo-911, were chosen from \cite{2014A&A...562A.146I}, where they have a metallicity in the range $-3\lesssim$ \feh $\lesssim-2$. The spectra of three more stars, Boo-094, Boo-117, and Boo-127, were reduced for comparison with other works.

The original spectra of the stars were obtained on the HDS echelle spectrograph \citep{2002PASJ...54..855N} with a resolution $R$ = 40\,000 and were taken by us from the Subaru archive (National Astronomical Observatory of Japan, proposal ID o09120, o10117). The spectrum of each star consists of several CCD frames with an exposure time from 1800 to 3600~s. The coordinates of the stars, their $V$ magnitudes (calculated from the SDSS $ugr$ magnitudes), and the characteristics of the observations are presented in Table~\ref{tab:obs}. The spectra were reduced in the MIDAS software in the echelle package. The following standard procedures were performed: bias subtraction, cosmic-ray hit removal, spectral order extraction, sky background subtraction, flat field correction, and normalization to the continuum level. We extracted 40 echelle orders on the blue CCD array in the wavelength range 3974--5450~\AA\ and 23 echelle orders on the red CCD array in the range 5385--6814~\AA. The typical $S/N$ ratio was 25--30 at the CCD sensitivity maximum ($\sim$6500~\AA). The spectral orders were combined by taking into account the noise in the overlapping parts of the spectrum. The radial velocity was estimated relative to the synthetic spectrum in the BinMag\footnote{http://www.astro.uu.se/\~{}oleg/binmag.html} code using a set of unblended lines.

\subsection{Measuring the equivalent widths and estimating their errors}

To measure the equivalent widths of spectral lines, we fitted them by the synthetic profile computed with preliminary stellar atmosphere parameters and fixed macroturbulent, $\xi_{macro}$=6~\kms, and rotational, $\varv_{rot}$ = 1~\kms, velocities typical for red giants. The elemental abundance and the radial velocity were free parameters. We then selected only the spectral line under study from the list of lines, computed its profile with the derived abundance, and calculated its equivalent width. The possible errors were estimated by two methods.

The first method is simple. The equivalent width of a spectral line is $EW=\int(1-f_\lambda)d\lambda$, where $f_\lambda$ is the normalized flux or, in the case of a CCD with a pixel size $\Delta\lambda$, the integral is replaced by the sum $\sum(1-f_\lambda)\Delta\lambda$. Let $SN=1/\delta f_\lambda$ be the ratio of the signal to the noise that is assumed to be a constant $\delta f_\lambda$ over the profile of a shallow line. The error in the equivalent width is then $\delta EW = \Delta\lambda\sqrt{n}/SN$, where $n$ is the number of pixels in the line profile. $SN$ is determined directly from the CCD frame. The typical $\delta EW$ for the spectra used is $\sim$10~m\AA.

The second method is more accurate. In the case where the profile of a spectral line is described by a Gaussian, we can use its properties. The equivalent width is expressed by the area of the Gaussian $EW=\sqrt{2\pi}f_{cen}\sigma$, where $f_{cen}$ is the central depth of the spectral line and  $\sigma$ is the Gaussian width. Then, $\delta\sigma=1.92\sigma\delta f_{cen}/f_{cen}$ from the derivative of the Gaussian function. Using the fact that the Gaussian dominates in the observed profile and the full width at half maximum of the line is expressed via the Gaussian width $FWHM=2.35\sigma$ and the resolution is $R=\lambda/0.849FWHM$ \citep{1992IUE........112W}, we obtain $\delta EW = 1.93\lambda\delta f/{R}$. In the line measurement program $\delta f_\lambda$ is calculated for each line as a standard deviation of the observed profile from the computed one:  $\delta f_\lambda=\sqrt{1/n\sum (f_\lambda-f^{synt}_\lambda)^2}$.  The error in the equivalent width is comparable to that in the previous case.

We separately took into account the error due to the uncertainty in drawing the continuum level $\delta f$, which is less than $1/SN$. Then, $\delta EW = EW\delta f/f_{cen}$. The total error in the equivalent width is from 15--25~m\AA\ at $SN$ = 20 ... 30 to 30--40~m\AA\ at $SN$ = 5 ... 10.

\subsubsection{Comparing the equivalent widths}

\begin{figure}
	\centering
	\includegraphics[width=1\columnwidth,clip]{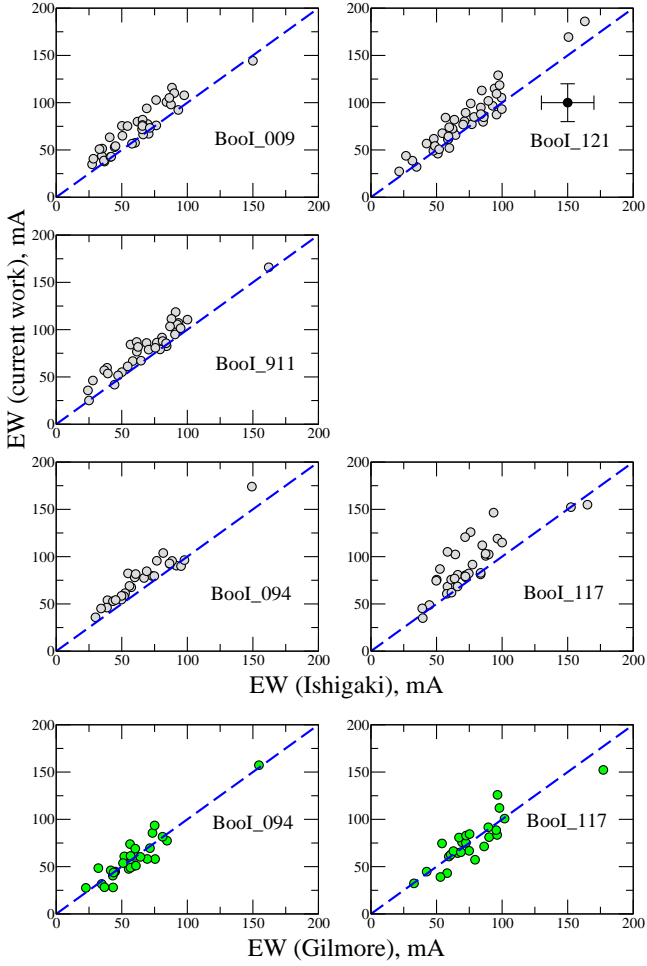}
	\caption{Comparison of the line equivalent widths between this paper and \citet{2014A&A...562A.146I} and \citet{2013ApJ...763...61G}. The bar of typical errors ($\sim$20~m\AA) is shown in the upper right corner.}
	\label{fig:EWcmp}
\end{figure}

\begin{figure}
	\centering
	\includegraphics[width=1\columnwidth,clip]{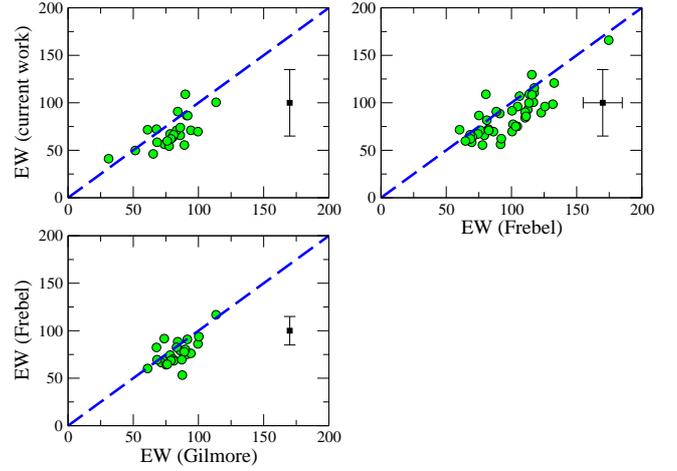}
	\caption{Comparison of the line equivalent widths in the spectra of Boo-127 taken with different instruments. The
		bar of typical errors ($\sim$20--30~m\AA) is shown in the upper right corner.}
	\label{fig:EWcmp2}
\end{figure}

The spectra of stars have a low S/N ratio, are significantly affected by the spectrum of scattered sky light, and the profiles of spectral lines are often distorted. Therefore, it is difficult to draw the continuum level. This raises the question about the possible presence of systematic errors in measuring the equivalent widths. For example, \cite{2014A&A...562A.146I} found their equivalent widths to be smaller than those measured by \cite{2013ApJ...763...61G} approximately by 18~m\AA.

We compared our measurements with the measurements of common stars in \citet[][UVES-FLAMES/VLT, $R=47\,000$]{2013ApJ...763...61G}, \citet[][HDS/Subaru, $R=40\,000$]{2014A&A...562A.146I}, and \citet[][Mike/Magellan, $R\approx28\,000$]{2016ApJ...826..110F}. These studies are based on the spectra taken with close resolutions, which ensures a proper comparison. The results for five common stars from the first paper and two from the second one are presented in Fig.~\ref{fig:EWcmp}. Our equivalent widths are seen to be systematically higher than those from \cite{2014A&A...562A.146I}, on average, by 11~m\AA\ or 15\%, although this difference is smaller than the measurement error. At the same time, the measurements by \cite{2013ApJ...763...61G} are very close to ours, there are no systematic differences, the mean difference is less than 1~m\AA. This means that the equivalent widths in \cite{2014A&A...562A.146I} were systematically underestimated.

We also checked the quality of our measurements (HDS/Subaru) using another star, Boo-127, which is common to \citet[][Mike/Magellan]{2016ApJ...826..110F} and \citet[][UVES/VLT]{2013ApJ...763...61G}, as an example. The HDS/Subaru observations were obtained on two nights, May 15--16, 2010 (proposal ID o10117). The reduction of CCD images and the method of equivalent width measurements are identical to those in this paper. The results are shown in Fig.~\ref{fig:EWcmp2}. Our measurements for this star are lower than those in the two other papers approximately by 10~m\AA, but within the error limits. The equivalent widths from \citet{2016ApJ...826..110F} are also slightly smaller, by 6~m\AA, than those from \cite{2013ApJ...763...61G}, but the differences do not exceed the typical errors either.

Thus, the accuracy of our equivalent width measurements is comparable to that in \cite{2013ApJ...763...61G} and \citet{2016ApJ...826..110F}.

\section{Stellar atmosphere parameters}
\label{Sect:Param}

\begin{table}
	\centering
	\caption{The effective temperatures of the investigated stars determined from various color indices.}
	\label{tab:T} 
	\renewcommand{\tabcolsep}{4pt}
	\begin{tabular}{cccccccc}
		\hline\hline
		Star&   \multicolumn{6}{c}{\Teff, K}\\
		&   $(B-V)$  &  $(V-R)$  &   $(V-I)$  &   $(V-J)$  &   $(V-H)$  &   $(V-K)$        \\
		\hline
		009  & 4531$\pm$20 & 4592$\pm$40 & 4484$\pm$24 & 4462$\pm$48 & 4503$\pm$60 & 4361$\pm$66 \\
		121  & 4531$\pm$20 & 4573$\pm$39 & 4467$\pm$22 & 4419$\pm$44 & 4466$\pm$60 & 4440$\pm$74 \\
		911  & 4547$\pm$21 & 4578$\pm$41 & 4456$\pm$24 & 4462$\pm$51 & 4445$\pm$55 & 4532$\pm$94 \\
		\hline
	\end{tabular}
\end{table}
	
To retain the homogeneity of the studies, we determined the effective temperatures \Teff of the stars in much the same way as was done by \paperI, from the $B-V$, $V-R$, $V-I$, $V-J$, $V-H$, and $V-K$ colors (see Table~\ref{tab:T}) using the calibrations by \cite{2005ApJ...626..465R}. The metallicities needed for our calculations were taken from \cite{2014A&A...562A.146I}. The $BVRI$ magnitudes were obtained by converting the SDSS DR12 $ugriz$ colors \citep{2015ApJS..219...12A} using the formulas from \cite{2006A&A...460..339J}. The infrared $JHK$ magnitudes were taken from the 2MASS catalogue \citep{2006AJ....131.1163S}. The interstellar reddening was assumed to be $E(B-V)=0.02$ \citep{2006ApJ...647L.111B} and was taken into account in all color indices. The final value of \Teff in Table~\ref{tab:param} is a weighted mean from all colors including the errors.

The surface gravity log g was calculated from the relation
\begin{eqnarray*}
	\mathrm{log}\,g&=&-10.607+\mathrm{log}\,M+4\,\mathrm{log}\,T_{eff}-\\
	& &-0.4\,[4.74-(m_V+BC+|m_V-M_V|+A_V)]
\end{eqnarray*}

where $M=0.8M_{\sun}$ is the stellar mass, $m_V$ is the $V$ magnitude, $BC$ is the bolometric correction calculated as prescribed by \cite{1999A&AS..140..261A}, $|m-M|=18.9^m$ is the distance modulus from \citet{2006ApJ...647L.111B}, and $A_V=3.1E(B-V)=0.06$~mag is the interstellar extinction in the $V$ band. The total error in \lgg$\approx$0.10~dex. Note that for old objects, such as stars in \boo, the mass of a star is reliably fixed, with an accuracy of 0.05~dex, if its evolutionary status has been established. In the galaxy \boo only giants can be observed with a high spectral resolution.

The microturbulence $\xi_t$ was estimated from the empirical formula derived for metal-poor Galactic halo giants by \cite{2017A&A...604A.129M}:
$$\xi_t = 0.14 - 0.08\, {\rm \feh} + 4.90\, (T_{eff} / 10^4) - 0.47\, {\mathrm{log}~g}$$
It provides an accuracy of about 0.2~\kms.

\begin{table}
	\centering
	\renewcommand{\tabcolsep}{4pt}
	\caption{Atmospheric parameters of the investigated stars and the difference in abundance between two ionization stages
		for iron and titanium} 
	\label{tab:param} 
	\begin{tabular}{ccccccc}
		\hline\hline
		Star& \Teff       & \lgg          &$\xi_t$& \feh &\ion{Fe}{1}-\ion{Fe}{2}   &  \ion{Ti}{1}-\ion{Ti}{2}\\
		&  K          &               &\kms &&                &  \\
		\hline      
		009 & 4500$\pm$70 & 1.22$\pm$0.10 & 2.0 &-3.14& ~0.13$\pm$0.29 & ~0.31$\pm$0.59 \\
		121 & 4490$\pm$70 & 1.21$\pm$0.10 & 2.0 &-2.73& ~0.08$\pm$0.18 & ~0.00$\pm$0.29 \\
		911 & 4500$\pm$70 & 1.24$\pm$0.10 & 2.0 &-2.32& -0.14$\pm$0.20 & -0.15$\pm$0.21 \\
		\hline
	\end{tabular}
\end{table}

Since the photometric data for the three new stars are close, we also expect to obtain close stellar atmosphere parameters. Table~\ref{tab:param} gives the estimated stellar atmosphere parameters. The estimated effective temperatures are close to those from \cite{2014A&A...562A.146I}, with the exception of Boo-009 for which they give \Teff$=4750$~K, while the surface gravities differ due to the application of different methods for determining \lgg. \cite{2014A&A...562A.146I} used the theoretical Yonsei-Yale (YY) isochrones with an age of 12~Gyr, while we follow the method described above and based on the distance modulus and photometry. The differences in microturbulence are insignificant, with the exception of Boo-009 for which \cite{2014A&A...562A.146I} give $\xi_t$ = 2.5~\kms. This may be a consequence of the difference in \Teff and \lgg and large errors in determining the iron abundance from spectra with a low $S/N$ ratio. 

To check the deduced stellar atmosphere parameters, we analyzed the \ion{Fe}{1}/\ion{Fe}{2} and \ion{Ti}{1}/\ion{Ti}{2} ionization equilibrium as well the positions of the stars on the HR diagram. The differences in abundance between two ionization stages are presented for iron and titanium in Table~\ref{tab:param}. In all cases, the difference does not exceed the measurement error. Thus, our analysis of the spectra confirms the atmospheric parameters estimated by non-spectroscopic methods.

\begin{figure}
	\centering
	\includegraphics[width=1\columnwidth,clip]{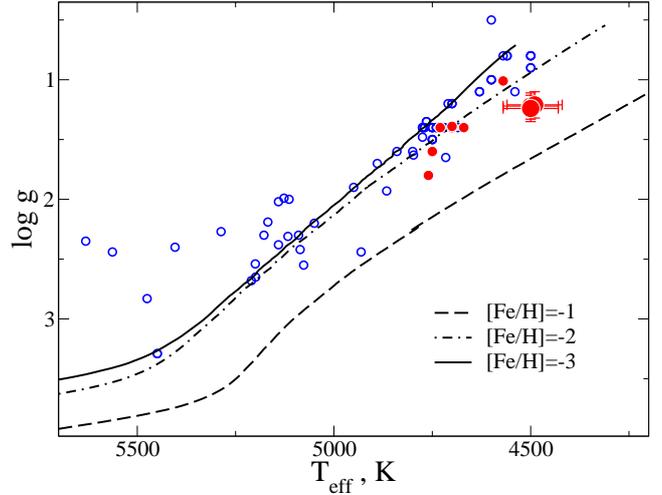}
	\caption{The positions of \boo stars on the HR diagram with evolutionary tracks for a mass of $0.8\,M_{\sun}$, metallicities \feh = $-1, -2, -3$ and \afe = 0.3. The big circles are the stars from this paper, while the small filled circles are the stars from \paperI. The error bar is shown for $1\sigma$. The published data are marked by the open circles.}
	\label{fig:HR}
\end{figure}

The positions of all \boo giants from \cite{2017A&A...604A.129M} and this paper are indicated by the filled circles in Fig.~\ref{fig:HR}, along with three evolutionary tracks constructed for a mass of $0.8\,M_{\sun}$ and metallicities\feh = $-1, -2, -3$ and \afe = 0.3 and obtained by interpolating the Y2 tracks \citep[version~3][]{2009gcgg.book...33H}. The open circles represent the published data collected in \cite{2016ApJ...826..110F}. Note that one star can be presented several times, according to the data of different authors. All of the red giants (\lgg $<2$) are located between the evolutionary tracks with metallicities \feh $=-1..-3$. However, the stars investigated in this paper are cooler than most of the other ones. For them to be placed on the track of the corresponding metallicity, \lgg must be reduced to 0.8--0.9. If the distance modulus is assumed to be 19.07~mag \cite{2012ApJ...744...96O}, then \lgg will decrease only by 0.07~dex. In the literature there are reports about a similar inconsistency of the positions of giant stars on the HR diagram with their evolutionary tracks, for example, for the halo giant HD~122563 with reliably measured atmospheric parameters \citep{2019A&A...625A..33C}. This may be related to uncertainties in the calculations of metal-poor tracks in the region of red giants.

\begin{table*}
	\centering
	\caption{The abundances of chemical elements in the atmospheres of the investigated stars.} 
	\label{tab:abund} 
	\begin{tabular}{lcrrrrrr}
		\hline
		Ion &$\eps{X_{\sun}}$& N& \multicolumn{3}{c}{LTE}& \multicolumn{2}{c}{non-LTE} \\
		(X)&& & $\eps{X}$  & [X/H]&[X/Fe]&   $\eps{X}$  &  [X/Fe]\\
		\hline
		\multicolumn{8}{c}{Boo-009}\\
		\ion{Na}{1} &6.33&  2&  3.30$\pm$0.01& -3.03& -0.02&    2.96$\pm$0.04& -0.36\\
		\ion{Mg}{1} &7.58&  3&  4.81$\pm$0.34& -2.77&  0.24&    4.75$\pm$0.38&  0.18\\
		\ion{Ca}{1} &6.36&  5&  3.62$\pm$0.15& -2.74&  0.27&    3.72$\pm$0.13&  0.37\\
		\ion{Ti}{1} &4.93&  2&  1.83$\pm$0.58& -3.10& -0.09&    1.84$\pm$0.58& -0.08\\
		\ion{Ti}{2} &4.93&  4&  1.91$\pm$0.09& -3.02& -0.01&    2.15$\pm$0.08&  0.23\\
		\ion{Fe}{1} &7.50&  3&  4.36$\pm$0.25& -3.14& -0.13&    4.36$\pm$0.25& -0.13\\
		\ion{Fe}{2} &7.50& 19&  4.35$\pm$0.14& -3.15& -0.14&    4.49$\pm$0.13&  0.00\\
		\ion{Ni}{1} &6.25&  2&  3.44$\pm$0.32& -2.81&  0.20&   -9.99$\pm$0.00&  0.33\\
		\ion{Ba}{2} &2.13&  2& -1.92$\pm$0.04& -4.05& -1.04&   -1.77$\pm$0.01& -0.89\\
		\multicolumn{8}{c}{Boo-121}                                           \\
		\ion{Na}{1} &6.33&  2&  3.45$\pm$0.06& -2.88& -0.15&    3.14$\pm$0.06& -0.46\\
		\ion{Mg}{1} &7.58&  4&  5.05$\pm$0.18& -2.53&  0.20&    5.05$\pm$0.13&  0.20\\
		\ion{Ca}{1} &6.36& 12&  3.92$\pm$0.09& -2.44&  0.29&    4.04$\pm$0.06&  0.41\\
		\ion{Ti}{1} &4.93& 11&  2.25$\pm$0.12& -2.68&  0.05&    2.49$\pm$0.13&  0.29\\
		\ion{Ti}{2} &4.93& 16&  2.49$\pm$0.26& -2.44&  0.29&    2.49$\pm$0.26&  0.29\\
		\ion{Fe}{1} &7.50& 41&  4.74$\pm$0.17& -2.76& -0.03&    4.85$\pm$0.17&  0.08\\
		\ion{Fe}{2} &7.50&  5&  4.77$\pm$0.07& -2.73&  0.00&    4.77$\pm$0.07&  0.00\\
		\ion{Ni}{1} &6.25&  4&  3.56$\pm$0.16& -2.69&  0.04&   -9.99$\pm$0.00&  0.07\\
		\ion{Ba}{2} &2.13&  4& -1.03$\pm$0.21& -3.16& -0.43&   -1.04$\pm$0.22& -0.44\\
		\multicolumn{8}{c}{Boo-911}                                           \\
		\ion{Na}{1} &6.33&  2&  3.47$\pm$0.06& -2.86& -0.54&    3.19$\pm$0.05& -0.82\\
		\ion{Mg}{1} &7.58&  2&  5.11$\pm$0.45& -2.47& -0.15&    5.11$\pm$0.35& -0.15\\
		\ion{Ca}{1} &6.36&  6&  3.93$\pm$0.17& -2.43& -0.11&    4.04$\pm$0.17&  0.00\\
		\ion{Ti}{1} &4.93& 10&  2.19$\pm$0.12& -2.74& -0.42&    2.41$\pm$0.11& -0.20\\
		\ion{Ti}{2} &4.93& 10&  2.55$\pm$0.18& -2.38& -0.06&    2.56$\pm$0.18& -0.05\\
		\ion{Fe}{1} &7.50& 33&  4.95$\pm$0.17& -2.55& -0.23&    5.04$\pm$0.17& -0.14\\
		\ion{Fe}{2} &7.50&  4&  5.18$\pm$0.10& -2.32&  0.00&    5.18$\pm$0.10&  0.00\\
		\ion{Ni}{1} &6.25&  2&  3.71$\pm$0.21& -2.54& -0.22&   -9.99$\pm$0.00&  0.01\\
		\ion{Ba}{2} &2.13&  3& -0.67$\pm$0.17& -2.80& -0.48&   -0.73$\pm$0.19& -0.54\\
		\hline
	\end{tabular} 
\end{table*}

\section{Determining the abundances of chemical elements}
\label{Sect:Abund}

Following the technique from our previous papers \citep{2017A&A...604A.129M, 2017A&A...608A..89M}, we use spherically symmetric model stellar atmospheres with standard abundances of chemical elements interpolated for specified \Teff/\lgg/\feh in the MARCS3 grid of models\footnote{http://marcs.astro.uu.se} \citep{2008A&A...486..951G}, the WIDTH9 code \citep[][modified by V.Tsymbal]{2005MSAIS...8...14K} to calculate the abundances from the equivalent widths, the SynthV \citep{2019ASPC....R} and SIU \citep{reetz} codes to compute the synthetic spectrum, the DETAIL code \citep{detail} to calculate the non-LTE populations of atomic levels, and the Vienna Atomic Line Database \citep{2015PhyS...90e4005R}.

The carbon abundance was determined under the LTE assumption from an analysis of the molecular line profiles by the synthetic spectrum method. The non-LTE approach was used to analyze the \ion{Na}{1}, \ion{Mg}{1}, and \ion{Ba}{2} line profiles. The \ion{Ba}{2} lines were computed by taking into account the hyperfine structure.

For the \ion{Ca}{1}, \ion{Ti}{1}, \ion{Ti}{2}, and \ion{Fe}{1} lines we used the method of equivalent widths by applying the non-LTE abundance corrections from \cite{2016AstL...42..606M}. The equivalent widths were measured by describing the observed profile by the synthetic one with fixed macroturbulence ($\varv_{macro}=6$~\kms) and rotation ($\varv_{rot} \sin i = 1$~\kms) velocities. The list of spectral lines used and their parameters were published in \paperII (hereafter MJS17). The derived LTE and non-LTE abundances of chemical elements ($\eps{X}$, [X/H] [X/Fe]) are presented in Table~\ref{tab:abund}, where the corresponding abundances from \cite{1989GeCoA..53..197A}, \citet[][Ti]{2003ApJ...591.1220L}, and \cite{1998SSRv...85..161G} are also given. We use the scale where $\eps{H}$ = 12 for hydrogen.

\begin{figure}
	\centering
	\includegraphics[width=1\columnwidth,clip]{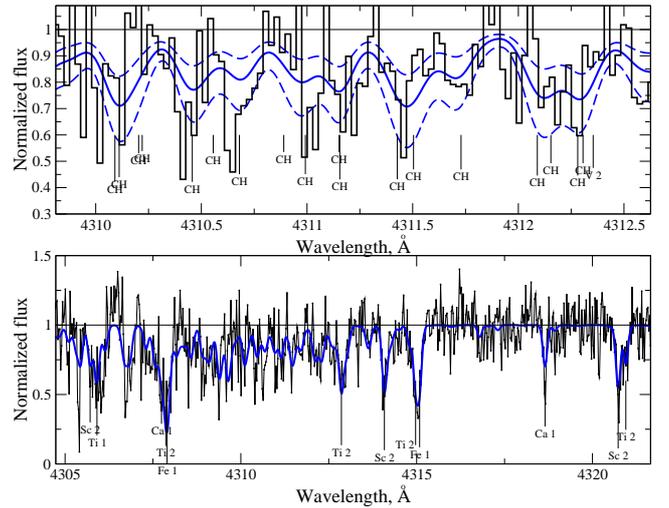}
	\caption{Top (a): Molecular CH lines in the spectrum of the star Boo-121. The size of the vertical bars reflects the contribution. The synthetic spectrum that describes the observations best is indicated by the thick solid curve. The dashed curves correspond to a variation in the carbon abundance of $\pm$0.30~dex. Borttom (b): Comparison of the observed and synthetic spectra near the CH band.}
	\label{fig:CH}
\end{figure}

\begin{table}
	\centering
	\caption{The carbon abundance $\eps{C}$ in the atmospheres of the investigated stars determined from molecular CH lines: 1)~4309.95--4312.45~\AA\ 2)~4313.43--4313.80~\AA\ 3)~4355.51--4356.75~\AA\ 4)~4363.93--4364.37~\AA.} 
	\label{tab:carbon} 
	\begin{tabular}{cccccc}
		\hline\hline
		Star& 1 & 2 & 3 & 4 & [C/Fe]\\
		\hline
		009 & 4.63$\pm$0.08 &  &  &  & -0.66 \\
		094 & 5.17$\pm$0.09 &  &  &  & -0.57 \\
		117 & 5.44$\pm$0.10 &  &  &  & -0.90 \\
		121 & 4.90$\pm$0.08 & 4.73$\pm$0.16 & 4.91$\pm$0.16 & 4.88$\pm$0.17 & -0.84 \\
		911 & 4.94$\pm$0.09 &  &  &  & -1.14 \\
		\hline
	\end{tabular}
\end{table}

\begin{figure}
	\centering
	\includegraphics[width=1\columnwidth,clip]{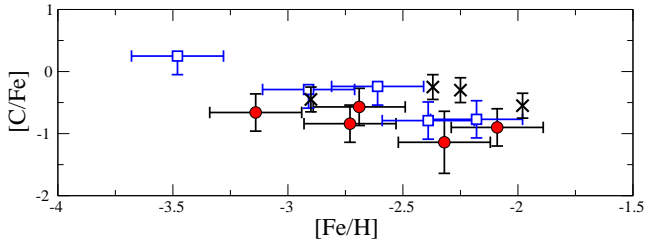}
	\caption{[C/Fe]abundance in the atmospheres of the investigated stars (filled circles) in comparison with the data from \citet{2014A&A...562A.146I} (open squares, the absence of an upper error bar means an upper limit) and \citet{2010ApJ...723.1632N} (crosses).
	}
	\label{fig:C-Fe}
\end{figure}

\subsection{Carbon}

Apart from Boo-009, Boo-121, and Boo-911, the carbon abundance was also determined for Boo-094 and Boo-117 using the atmospheric parameters from \cite{2017A&A...604A.129M}: \Teff/\lgg/\feh/$\xi_t$ = 4570/1.01/$-2.69$/2.2 and 4700/1.40/$-2.09$/2.3, respectively. Unfortunately, for metal-poor stars the carbon abundance can be estimated only from an analysis of the molecular CH bands located in the blue part of the spectrum with a high noise level. We used the molecular CH $A^2\Delta-X^2\Pi$~(0,0) band in the region $\lambda$4309.95--4312.45~\AA. Here the molecular lines dominate and remain  observable even in metal-poor stars at $S/N\lesssim 10$ (Fig.~\ref{fig:CH}). In the least noisy spectrum of Boo-121 we were able to also measure the molecular lines in the regions 4313.43--4313.80, 4355.51--4356.75, and 4363.93--4364.37~\AA\ and to calculate the weighted mean carbon abundance. The results are presented in Table~\ref{tab:carbon}. To calculate [C/Fe], we used the solar abundances of carbon $\eps{C}=8.43$ \citep{2015MNRAS.453.1619A} and iron $\eps{Fe}=7.50$. The abundance errors reflect the error of the method, while the total error, including the noise and the uncertainties in drawing the continuum level, is about 0.30~dex and reaches 0.50~dex for Boo-911. The results can be improved when using higher-quality spectra ($S/N > 20$). Since the main source of uncertainties is the sky background on the CCD array, whose level for faint stars exceeds considerably the useful signal, we analyzed the quality of its allowance in this region. We checked the continuum level reconstructed from the flat field correction. The nearest \ion{Ca}{1}, \ion{Fe}{1}, \ion{Fe}{2}, \ion{Ti}{1}, \ion{Ti}{2}, and \ion{Sc}{2} lines in the observed spectrum are well described by the synthetic spectrum with a specified abundance (see Fig.~\ref{fig:CH}). 

In Fig.~\ref{fig:C-Fe} we compare our values of [C/Fe] with those determined by \citet{2014A&A...562A.146I} from the same spectra and \citet{2010ApJ...723.1632N} for the same stars. Low-resolution ($R = 5\,000$) spectra were used in the latter paper. \citet{2014A&A...562A.146I} and \citet{2010ApJ...723.1632N} obtained a systematically higher carbon abundance, by 0.5--1 dex for different stars. According to our data, all stars in \boo have a carbon underabundance relative to iron with [C/Fe]$<-0.5$, which is typical for high-luminosity metal-poor stars \citep[see Fig.~7 in][]{2015A&A...583A..67J}. For comparison, for stars with similar parameters in the dwarf galaxy Tucana~III \citet{2018arXiv181201022M} obtained [C/Fe]$\approx-0.5$.

\subsection{Barium}
\label{sect:ba_abund}

In this paper we determined the abundance of barium for the three new stars added to the \boo sample and redetermined its abundance in the eight stars investigated by \cite{2017A&A...608A..89M}. The publication of quantum-mechanical calculations of \ion{Ba}{2} + \ion{H}{1} collisions \citep{2018MNRAS.478.3952B} necessitates a revision. In the atmospheres of metal-poor stars the electron density is low and the inelastic processes during collisions with neutral hydrogen atoms play an important role in establishing the statistical equilibrium of atoms. Due to the absence of accurate data, the formula from \cite{1984A&A...130..319S} based on the theoretical approximation of \cite{1968ZPhy..211..404D} has been used for a long time to calculate the rates of collisions with HI atoms. It has been established empirically that the Drawin \ion{Ba}{2} + \ion{H}{1} collision rates require scaling with a factor \kH\, = 0.01 \citep{1999A&A...343..519M}. The scaled Drawin rates were used by MJS17 in their non-LTE calculations for \ion{Ba}{2}. 

\begin{figure}
	\includegraphics[width=1\columnwidth, clip]{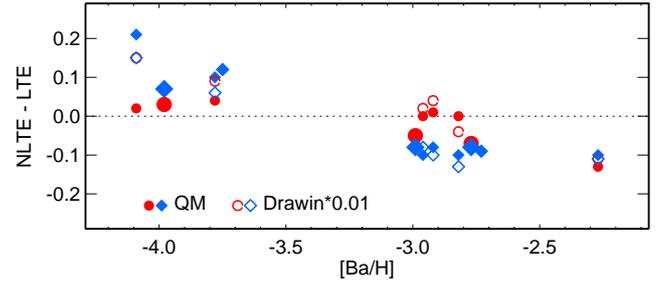}
	\caption{The non-LTE corrections for the \ion{Ba}{2} 6497~\AA\ (circles) and \ion{Ba}{2} 4934~\AA\ (diamond) lines in the \boo stars calculated using the quantum-mechanical (filled symbols) and Drawing (open symbols)  \ion{Ba}{2} + \ion{H}{1} collision rates. The bigger symbols correspond to the three stars for which the non-LTE barium abundance has been determined for the first time (only with accurate collisional data).}
	\label{fig:dnlte}
\end{figure}

As was first shown by \cite{1999A&A...343..519M}, the departures from LTE for \ion{Ba}{2} lines can be of different sign and magnitude, depending on the star's metallicity and its \Teff and \lgg. For our sample non-LTE leads to a strengthening of the \ion{Ba}{2} 4554, 4934~\AA\ resonance lines and negative abundance corrections, $\Delta_{\rm NLTE} = \eps{NLTE} - \eps{LTE}$ at [Ba/H] $> -3$, but to an opposite effect at a lower barium abundance (Fig.~\ref{fig:dnlte} for \ion{Ba}{2} 4934~\AA). For the subordinate \ion{Ba}{2} 5853, 6141, 6497~\AA\ lines the transition from negative to positive corrections occurs at a slightly higher barium abundance (Fig.~\ref{fig:dnlte} for \ion{Ba}{2} 6497~\AA). In comparison with the scaled Drawin rates, using accurate collisional data \citep{2018MNRAS.478.3952B} leads to a decrease in the departures from LTE at [Ba/H] $> -3$. The same is true for \ion{Ba}{2} 6141, 6497~\AA\ and at a lower barium abundance, but the non-LTE effects are enhanced for the \ion{Ba}{2} resonance lines. 

\begin{table}
	\caption{Revised non-LTE barium abundance in the \boo stars. The atmospheric parameters and LTE abundances are the same as those in \paperII.}
	\label{tab_ba}
	\tabcolsep 4pt
	\begin{center}
		\begin{tabular}{rccccccc}
			\hline\hline
			Star & \Teff & \lgg & \feh & $\xi_t$  & N & \multicolumn{2}{c}{$\eps{Ba}$} \smallskip \\ 
			\cline{7-8}  \noalign{\smallskip}
			& K &   &  & \kms  &  &  LTE & non-LTE \\
			\hline
			33 & 4730 & 1.4 & $-$2.35 & 2.3 & 3 & $-$0.73$\pm$0.15 & $-$0.70$\pm$0.17 \\
			41 & 4750 & 1.6 & $-$1.60 & 2.0 & 2 & $-$0.02$\pm$0.15 & $-$0.14$\pm$0.16 \\
			94 & 4570 & 1.01 & $-$2.90 & 2.2 & 2 & $-$1.74$\pm$0.40 & $-$1.66$\pm$0.47 \\
			117 & 4700 & 1.4 & $-$2.30 & 2.3 & 3 & $-$0.85$\pm$0.42 & $-$0.80$\pm$0.46 \\
			127 & 4670 & 1.4 & $-$2.10 & 2.3 & 2 & $-$0.62$\pm$0.02 & $-$0.60$\pm$0.01 \\
			130 & 4730 & 1.4 & $-$2.35 & 2.3 & 3 & $-$0.87$\pm$0.42 & $-$0.83$\pm$0.45 \\
			980 & 4760 & 1.8 & $-$3.00 & 1.8 & 2 & $-$1.77$\pm$0.01 & $-$1.63$\pm$0.03 \\
			1137 & 4700 & 1.39 & $-$3.7 & 1.9 & 3 & $-$2.07$\pm$0.11 & $-$1.98$\pm$0.04 \\
			\noalign{\smallskip}\hline \noalign{\smallskip}
		\end{tabular}
	\end{center}
\end{table}  %

For Boo-009, 121, and 911 we used all of the lines that we were able to measure in the spectrum, from 3 to 4 lines. When calculating the \ion{Ba}{2} 4554, 4934~\AA\ resonance lines, we took into account the fact that each of the five common isotopes $^{134}$Ba, $^{135}$Ba, $^{136}$Ba, $^{137}$Ba, and $^{138}$Ba forms its own line and, in addition, the lines of isotopes with an odd mass number consist of a set of hyperfine splitting components. For a solar mixture of isotopes the \ion{Ba}{2} 4554~\AA\ and \ion{Ba}{2} 4934~\AA\ lines have 15 and 11 components, respectively. Since we observe only old objects in \boo, where barium mostly likely originates exclusively in the r-process, we used the isotope abundance ratio predicted by \cite{1999ApJ...525..886A} for the r-process when calculating the relative intensity of the components: $^{135}$Ba : $^{137}$Ba : $^{138}$Ba = 26 : 20 : 54. The isotopes $^{134}$Ba and $^{136}$Ba are not formed in the r-process. The wavelengths and oscillator strengths of the components are the same as those in our previous papers \citep[see  e.g.][]{2014A&A...565A.123M}. The mean LTE and non-LTE abundances are presented in Table~\ref{tab:abund}. 

When determining the barium abundance in eight stars from our previous list, we used the same \ion{Ba}{2} lines and the same atomic line parameters as those in MJS17 (from Table 3). The results obtained are presented in Table~\ref{tab_ba}. Except for the star with the lowest metallicity, Boo-1137 ([Fe/H]$=-3.76$, [Ba/H]$=-4.09$), the changes in the barium abundance compared to MJS17 do not exceed 0.05~dex. For Boo-1137 \paperII used only the \ion{Ba}{2} 4934~\AA\ line. The subordinate \ion{Ba}{2} 6141 and 6497~\AA\ lines are weak, with the equivalent widths EW = 14.4 and 13.6~m\AA, respectively. The non-LTE abundance from the subordinate lines was found to be higher than that from the resonance line, by 0.2--0.3~dex, and this was explained by the EW measurement errors. However, applying the quantum-mechanical \ion{Ba}{2} + \ion{H}{1} collision rates led to a decrease in the non-LTE corrections for \ion{Ba}{2} 6141 and 6497~\AA\ and, conversely, to an increase in the non-LTE correction for \ion{Ba}{2} 4934~\AA, so the non-LTE abundances from different lines now agree within 0.09~dex.

\subsection{Uncertainties in the derived abundances}

\begin{table}
	\centering
	\caption{Uncertainties in the abundances of chemical elements for the star Boo-911 due to the errors in the 		atmospheric parameters. } 
	\label{tab:uncer} 
	\begin{tabular}{lcccc}
		\hline\hline
		Ion          & N  & \multicolumn{3}{c}{$\Delta \epsilon$}       \\
		\cline{3-5}\noalign{\smallskip}
		&    & \small $\Delta$\Teff = 100 K & \small $\Delta \xi_t=0.1$~\kms & Total\\
		&    & \small $\Delta$\lgg = 0.06   &           &\\
		\hline
		CH           & -  & 0.17 & 0.01 & 0.17 \\
		\ion{Na}{1}  & 2  & 0.20 & 0.11 & 0.23 \\
		\ion{Mg}{1}  & 3  & 0.08 & 0.04 & 0.09 \\
		\ion{Ca}{1}  & 6  & 0.10 & 0.04 & 0.11 \\
		\ion{Ti}{1}  & 10 & 0.19 & 0.03 & 0.19 \\
		\ion{Ti}{2}  & 10 & 0.04 & 0.06 & 0.07 \\
		\ion{Fe}{1}  & 33 & 0.17 & 0.08 & 0.19 \\
		\ion{Fe}{2}  & 4  & 0.00 & 0.04 & 0.04 \\
		\ion{Ni}{1}  & 2  & 0.12 & 0.03 & 0.12 \\
		\ion{Ba}{2}  & 3  & 0.08 & 0.12 & 0.14 \\
		\hline
	\end{tabular}
\end{table}

In the star Boo-911 for both iron and titanium the abundances from the lines of neutral atoms turn out to be lower, approximately by 0.15~dex, than those from the lines of ions. We analyzed the uncertainties in the abundance $\Delta \eps{X}$ due to the change in effective temperature $\Delta$\Teff\ = 100~K, surface gravity $\Delta$\lgg\ = 0.06 (the effect is the combined one, because \lgg depends on \Teff), and microturbulence  $\Delta \xi_t=0.1$~\kms\ as well as the total uncertainty. Our calculations were performed for $N$ spectral lines of each ion. For carbon we used the region 4309.95--4312.45~\AA, which contains more than 20 molecular CH lines. The results are presented in Table~\ref{tab:uncer}. We see that the uncertainties in \Teff and \lgg lead to greater uncertainties in the abundance from the \ion{Fe}{1} and \ion{Ti}{1} lines than from the \ion{Fe}{2} and \ion{Ti}{2}. Thus, the detected \ion{Fe}{1}-\ion{Fe}{2} and \ion{Ti}{1}-\ion{Ti}{2} differences fit into the range of possible uncertainties and may be attributable to the typical error in determining the atmospheric parameters.

\begin{figure}
	\centering 
	\includegraphics[width=1\columnwidth,clip]{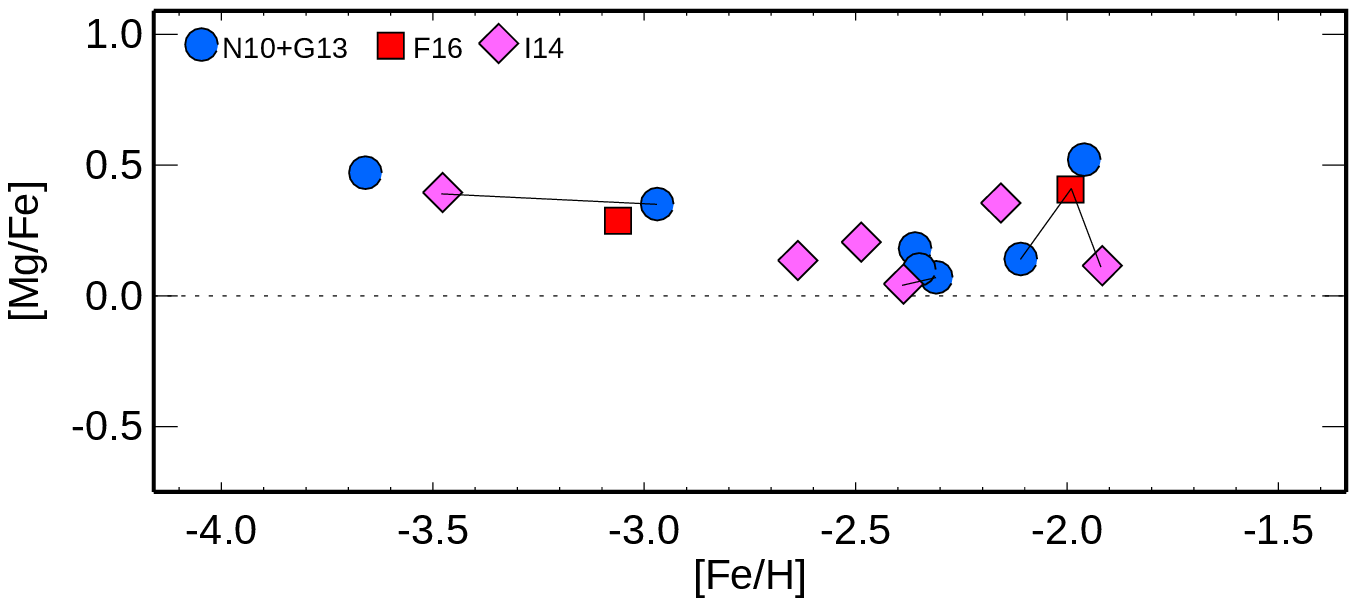}\\
	\includegraphics[width=1\columnwidth,clip]{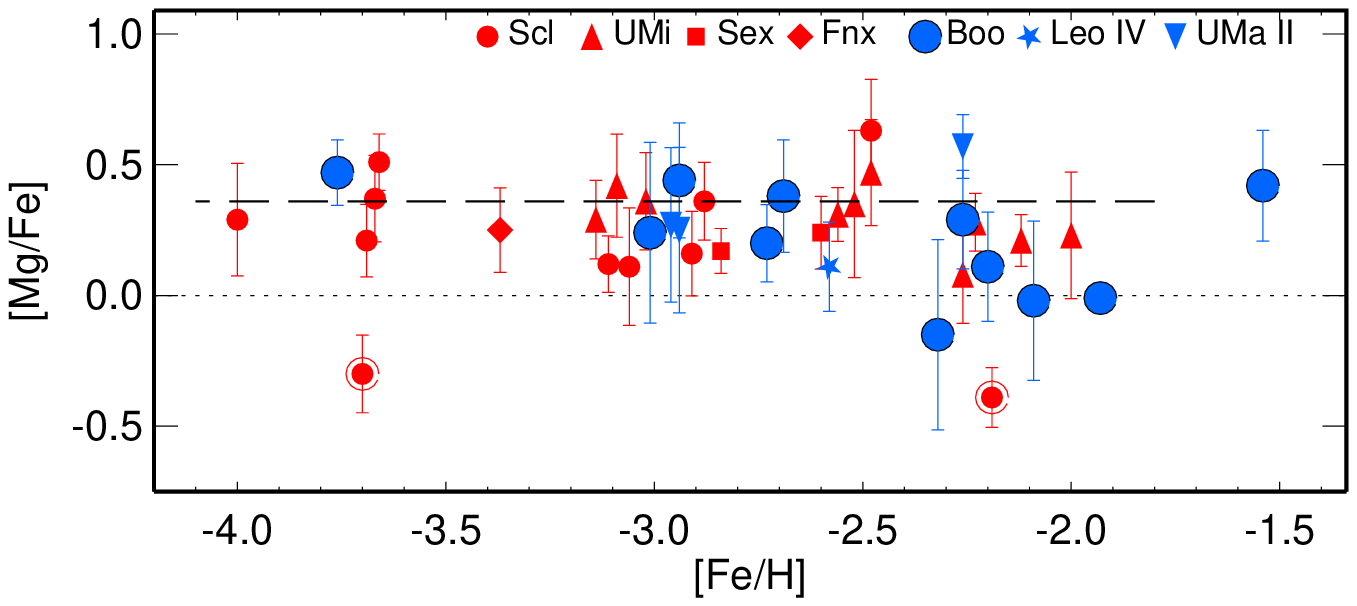}
	\caption{ Top (a): The [Mg/Fe] abundances in the \boo stars from \citet[][N10]{2010ApJ...723.1632N}, \citet[][G13]{2013ApJ...763...61G}, \citet[][I14]{2014A&A...562A.146I}, and \citet[][F16]{2016ApJ...826..110F}. The identical stars are connected by lines. Bottom (b): The [Mg/Fe] abundances from our (3 stars) and MJS17 data for other \boo stars and stars in other galaxies. The classical spheroidal galaxies and UFD galaxies are marked by the gray and black colors, respectively. The gray circle inside a bigger circle marks the stars with an underabundance of Mg and Ca relative to Fe. The dashed line marks the mean [Mg/Fe] ratio for Galactic halo stars from the MJS17 data.}
	\label{fig:Mg}
\end{figure}

\begin{figure}[t]
	\centering 
	\includegraphics[width=1\columnwidth,clip]{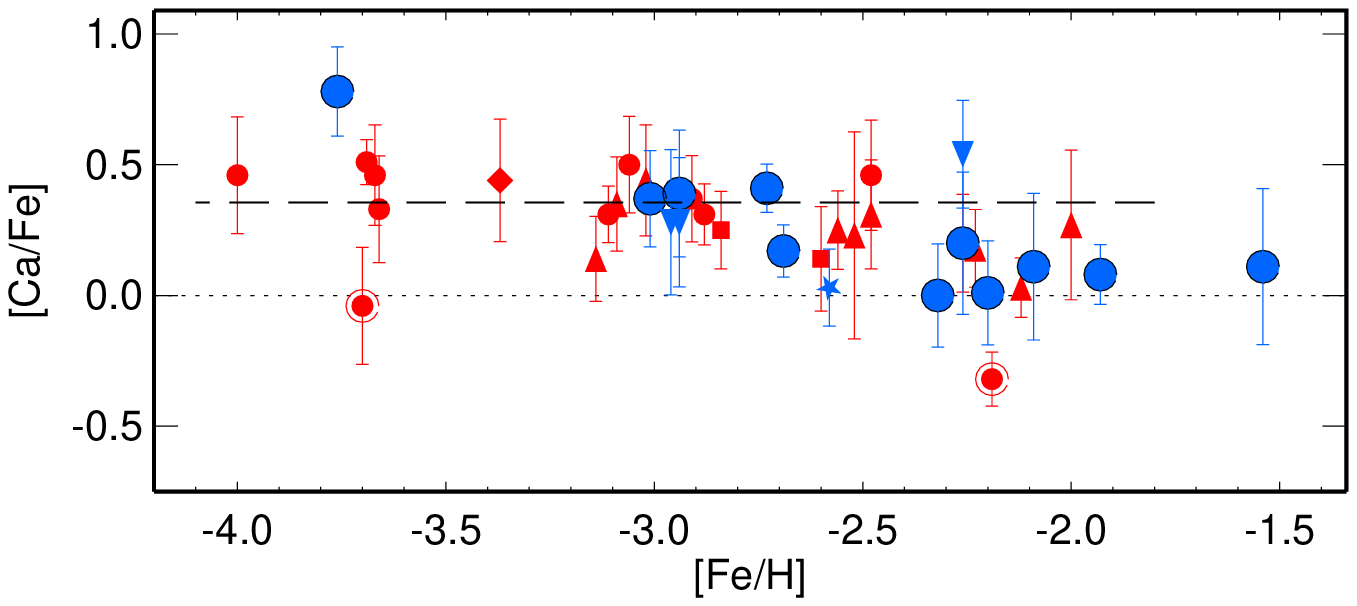}\\
	\includegraphics[width=1\columnwidth,clip]{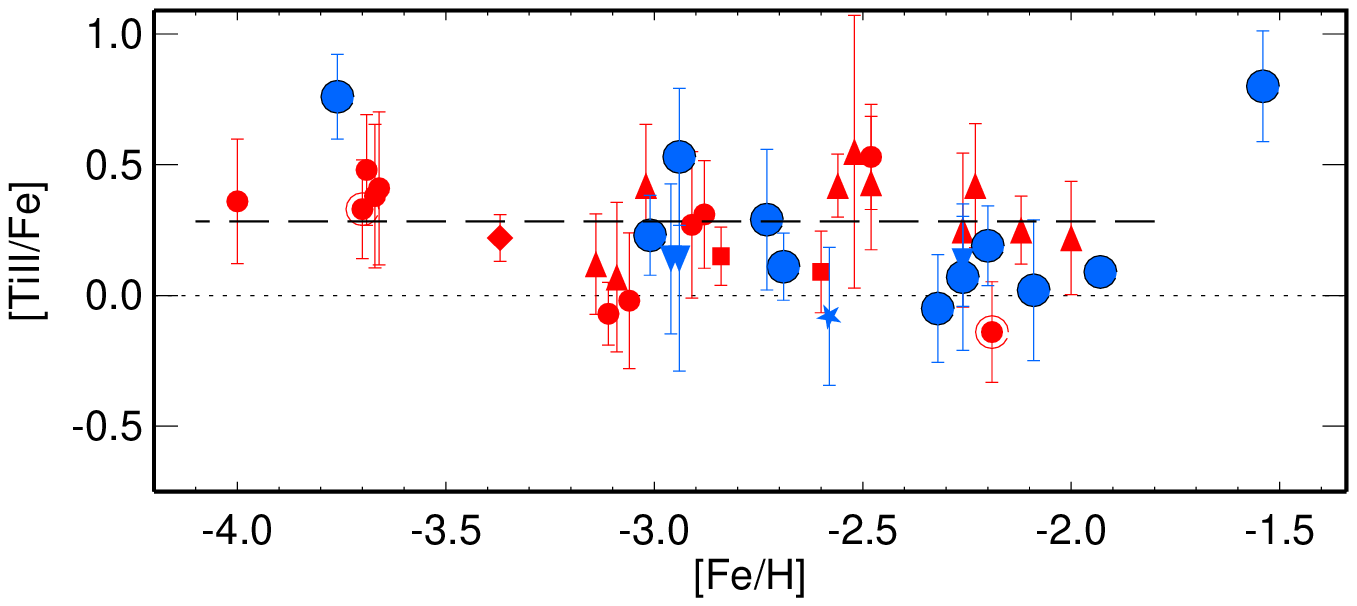}
	\caption{Same as Fig.~\ref{fig:Mg}b for [Ca/Fe] and [Ti/Fe].}
	\label{fig:alpha}
\end{figure}

\begin{figure}
	\centering 
	\includegraphics[width=1\columnwidth,clip]{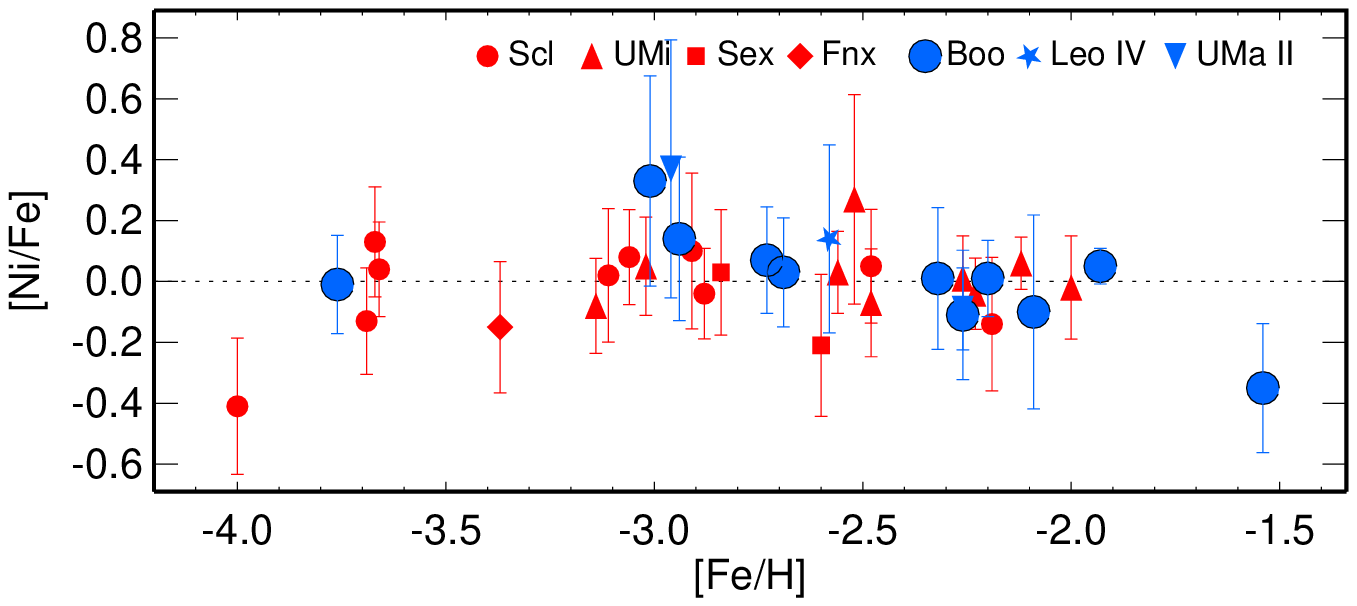}\\
	\includegraphics[width=1\columnwidth,clip]{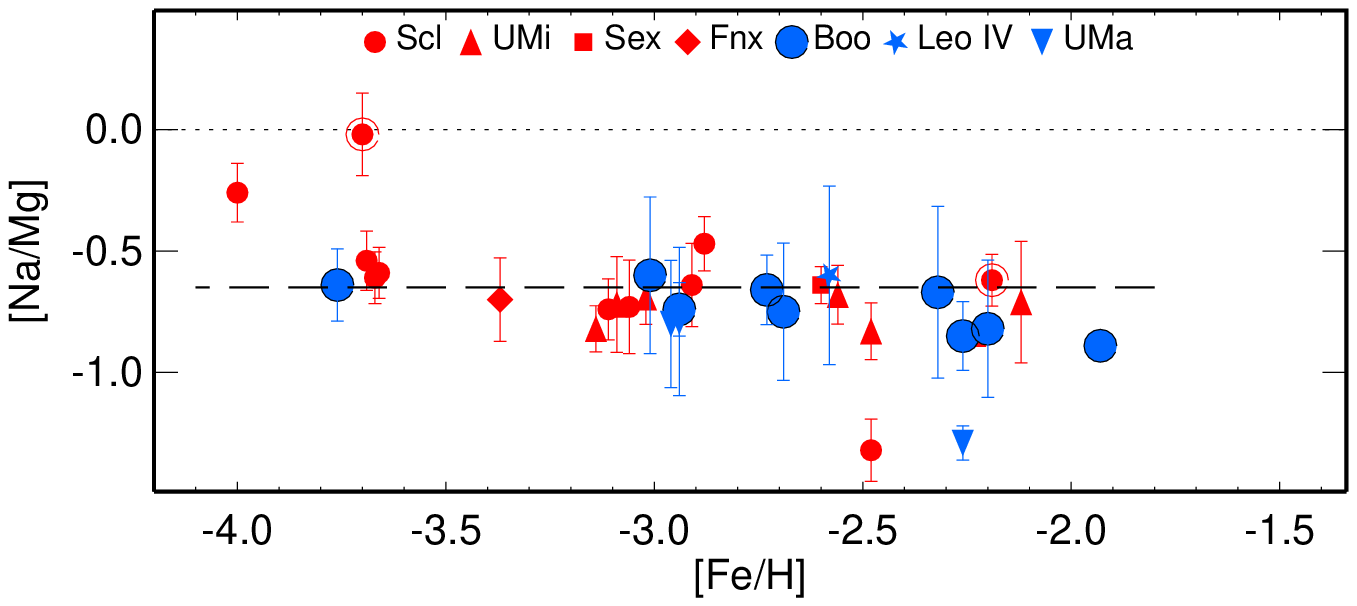}\\
	\includegraphics[width=1\columnwidth,clip]{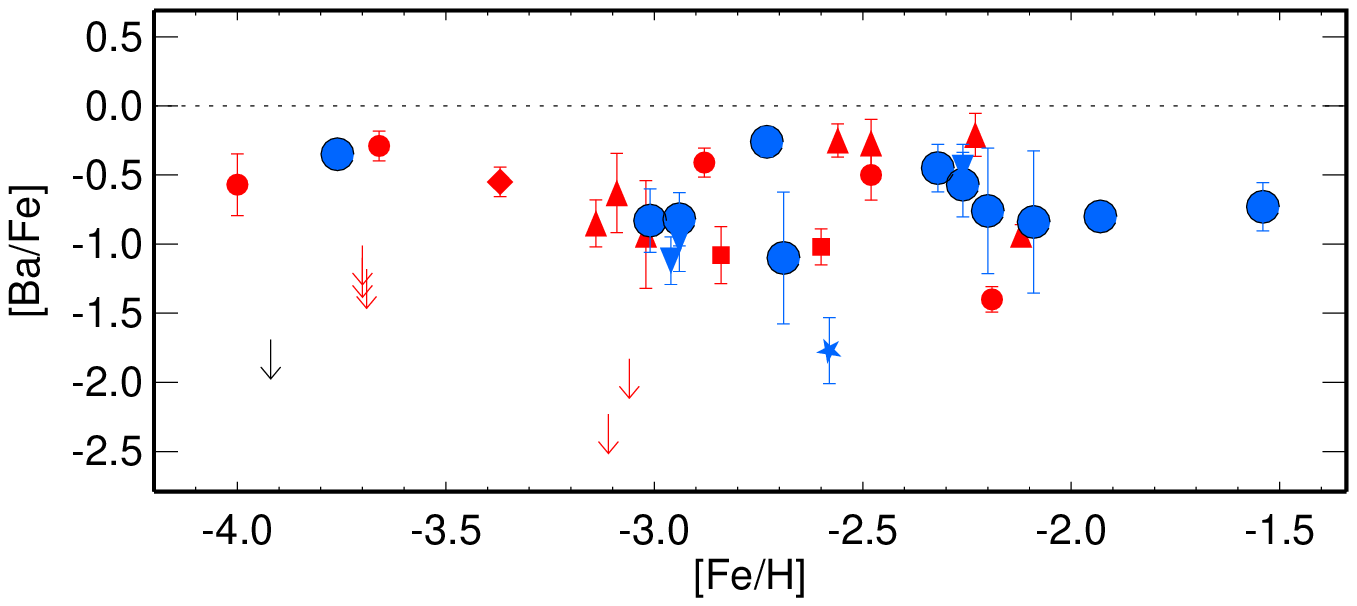}
	\caption{ Same as Fig.~\ref{fig:Mg}b for [Ni/Mg], [Na/Fe], and [Ba/Fe].}
	\label{fig:abund}
\end{figure}

\section{Discussion}
\label{Sect:discussion}

Following \paperI, we use the iron abundance derived from \ion{Fe}{2} line as a metallicity indicator, but it should be noted that when LTE is abandoned, the abundances from \ion{Fe}{1} and \ion{Fe}{2} lines coincide within the error limits.

\subsection{The \afe--\feh dependence}

We determined the abundances of three $\alpha$-process elements: Mg, Ca, and Ti. Figures~\ref{fig:Mg} and \ref{fig:alpha} show both our results and the MJS17 data for \boo and other dwarf galaxies for each of these elements. For a sample of Galactic halo comparison stars MJS17 obtained the same overabundance of each of the $\alpha$-process elements relative to iron, \afe$\approx0.3$. This is explained by the dominance of the explosions of massive SNe~II \cite[see, e.g.][]{1995MNRAS.277..945T}, in which the $\alpha$-elements are the main product and they are produced approximately twice as much as iron.

Observations show that for a galaxy of any type there is a range of low metallicities where \afe is constant, implying a decisive influence of SNe~II on the chemical evolution. The beginning of the decrease in \afe or the knee in the metallicity dependence, also called the $\alpha$-knee, points to the appearance of the first SNe~Ia in the galaxy and the increased production of iron. Determining the metallicity (or age) of the $\alpha$-knee (\feh knee ) is very important for understanding the star formation processes in a galaxy. In our Galaxy this knee is located at \feh knee $\approx -0.9$ \citep[see][and references therein]{2016ApJ...833..225Z}. In dwarf galaxies it is shifted toward lower metallicities, suggesting lower star formation rates. For example, according to \citet{2014MNRAS.443..658D}, \feh knee $\approx -1.3$ in the Sagittarius galaxy, \feh knee $\approx -1.7$ in the Large Magellanic Cloud, \feh knee $\approx -1.9$ in the Sculptor galaxy, and \feh knee $\approx -2.1$ in the Fornax galaxy.

It is very important to establish whether there is a knee in the \afe--\feh dependence for UFD galaxies and, if it is, then what \feh knee is. We have a sufficient number of investigated stars, 11, to study this problem only for the galaxy \boo. Since the value of \afe itself is comparatively small ($\sim$0.3), a very high accuracy of its determination is needed to detect a change. A simple compilation of published data does not allow definitive conclusions to be reached due to data inhomogeneity, as confirmed by Fig.~\ref{fig:Mg}a. Both individual authors and the entire data set show a decrease in [Mg/Fe] in the range $-3.7\le$ \feh $\le$ $-2.3$ from $\simeq 0.5$ to its solar value. On this basis, \citet{2013ApJ...763...61G} and \citet{2014A&A...562A.146I} hypothesized that SNe~Ia had managed to contribute to the chemical enrichment of \boo before the star formation stopped in the galaxy. However, in that case, it is unclear why [Mg/Fe] grows at \feh $>-2.3$, although this growth may be apparent and attributable to the abundance errors. For example, for one common star (Boo-127) the three papers give [Mg/Fe] differing almost by 0.4~dex.

Our [Mg/Fe] determinations are presented in the lower part of Fig.~\ref{fig:Mg}, along with the MJS17 data for other stars in \boo and stars in other dwarf galaxies. As was noted by MJS17, the star with the highest iron abundance, Boo-41 (\feh = $-1.54$), which is unusual per se for an UFD galaxy, has anomalously high magnesium and titanium abundances (Fig.~\ref{fig:alpha}) and an anomalously low nickel abundance (Fig.~\ref{fig:abund}). The chemical composition of this star apparently reflects a very rare type of nucleosynthesis episodes that do not affect the overall pattern of chemical evolution. We disregard Boo-41 when discussing the $\alpha$-trends. In the range $-3.8\lesssim$ \feh $\lesssim -2.7$ all five stars exhibit a Mg overabundance relative to Fe, [Mg/Fe]$\approx$0.3, while five stars with \feh from $-2.3$ to $-1.9$ have, on average, a solar Mg/Fe. Thus, we do not confirm the above-discussed growth of [Mg/Fe] in stars with \feh $>-2.3$. The two other $\alpha$-process elements, calcium and titanium, behave in the same way as does magnesium: in stars with \feh $\lesssim -2.7$ [Ca/Fe] $\approx$ [Ti/Fe] $\approx$ 0.3, while stars with \feh $\approx -2$ have solar Ca/Fe and Ti/Fe ratios. However, it should be noted that the star with the lowest iron abundance, Boo-1137, has an anomalously large overabundance of calcium and titanium.

The three stars added in this paper to the sample of \boo stars increased considerably the reliability of our conclusions about the behavior of \afe. In two of them \feh$<-2.7$ and both show \afe$\approx$0.3, while the third star with \feh $\approx -2.3$ has \afe$\approx$0.

Thus, it can be said with confidence that in the galaxy \boo in the range $-2.7\lesssim$ \feh $\lesssim -2.3$ there is a drop in \afe with increasing \feh. For any galaxy such a behavior is usually interpreted as the appearance of the first SNe~Ia and an increase in the iron production rate \citep[see, e.g.][]{1986A&A...154..279M}. The appearance of the first SNe a requires a star formation time scale of $\sim$1 Gyr, although there are theoretical works that predict a shorter time scale \citep[see, e.g.][$\approx$0.1~Gyr]{2012A&A...538A..82R}. Based on an analysis of the observed color--magnitude diagram, \citet{2012ApJ...744...96O} think that the stellar population of \boo was formed almost simultaneously. At the same time, modeling the chemical evolution \boo admits the possibility of an SNe~Ia contribution to the production of iron \citep{2015MNRAS.446.4220R}.

\subsection{Sodium, nickel, barium}

The [Ni/Fe], [Na/Mg], and [Ba/Fe] elemental ratios are presented in Fig.~\ref{fig:abund}. The stars in different galaxies, irrespective of their mass, have a nearly solar [Ni/Fe] ratio, suggesting a common origin of nickel and iron.

The stars in \boo exhibit close values of Na/Mg with its mean [Na/Mg] = $-0.73\pm0.12$. Other galaxies, including the Milky Way, also show the same plateau. This means that the sodium synthesis processes during carbon burning were identical in all systems, irrespective of their mass.

Our revision of the barium abundance and the inclusion of three new stars in the sample did not affect the MJS17 conclusions about the behavior of [Ba/Fe]. A barium underabundance relative iron, on average, with [Ba/Fe]$\approx -0.7$, is observed in \boo, although two stars with close metallicities (\feh $\approx -2.7$) show a difference in [Ba/Fe] by about an order of magnitude. Despite this,regard to the barium abundance the galaxy \boo is much more homogeneous than, for example,classical spheroidal galaxy in Sculptor, in which an upper limit on the barium abundance can estimated for 5 of the 11 stars. Before the appearance of the first intermediate-mass ($M = 2-4 M_\odot$) AGB stars and the onset of nucleosynthesis in the process, barium was synthesized in the r-process. \boo we definitely observe no growth of [Ba/Fe] with increasing \feh that could point to the onset of synthesis in the s-process.

This paper was motivated by the possibility not only to refine the $\alpha$-trends in \boo, but also to determine the strontium abundance in three more stars and, thus, to increase the number of stars in \boo with a known strontium abundance to six. \paperII raised the question about a difference in the behavior of [Sr/Ba] in massive galaxies (the Milky Way and classical dwarf spheroidal galaxies) and the smallest (UFD) galaxies. Unfortunately, in the spectra of the stars under study the range where the \ion{Sr}{2} 4215.5~\AA\ line is located has a low S/N ratio and the line cannot be reliably measured. New, higher-quality observations are needed to study the Sr/Ba problem in the galaxy \boo.

\section{Conclusions}
\label{Sect:conc}

Using the methods developed in our previous studies \citep{2017A&A...604A.129M, 2017A&A...608A..89M}, we enlarged the sample of stars in the galaxy \boo for which a homogeneous set of atmospheric parameters was obtained and the abundances of key chemical elements were determined from high-resolution spectra and without using the LTE assumption to 11. Thus, among the UFD galaxies \boo now has a maximum number of thoroughly studied stars. Using three $\alpha$-process elements (magnesium, calcium, and titanium), we confirmed the previously suspected change in the \afe--\feh trend, namely the transition from their overabundance [$\alpha$/Fe] $\simeq$ 0.3 to the solar $\alpha$/Fe ratio, and established that it takes place in the range $-2.7 \lesssim$ \feh $\lesssim -2.3$. This suggests a contribution of type Ia supernovae to the chemical enrichment of the galaxy \boo. With regard to the sodium, nickel, and barium abundances \boo does not differ from other galaxies, both UFD and massive ones, such as classical dwarf spheroidal galaxies and the Milky Way.

\section*{Acknowledgments}

This work was supported by Program No12 of the Presidium of the Russian Academy of Sciences ''Questions of the Origin and Evolution of the Universe Using Methods of Ground-Based Observations and Space Research''. The work presented in Section~\ref{sect:ba_abund} was performed as part of RSF project no. 17-13-01144.

\bibliographystyle{mnras}
\bibliography{paper}

\end{document}